# К вопросу об омичности контактов Шоттки

© *А.В. Саченко, А.Е. Беляев, Р.В. Конакова*

Институт физики полупроводников им. В. Е. Лашкарева Национальной академии наук Украины,

03028 Киев, Украина

Аннотация

Проанализированы условия реализации омических контактов в случае контактов Шоттки. На основе классических представлений о механизмах токопрохождения рассмотрена обобщенная модель контакта Шоттки, учитывающая термоэлектронный ток основных носителей заряда и рекомбинационный ток неосновных носителей заряда в контактах Шоттки с диэлектрическим зазором. Анализ результатов использованной модели позволил получить критерии омичности контактов Шоттки и сопоставить между собой условия малого уровня инжекции и омичности контактов Шоттки для контактов на основе кремния. Показано, что условия омичности контакта Шоттки не совпадают с таковыми для случая $p-n$ переходов.

## On the issue of ohmicity of Schottky contacts

© *A.V. Sachenko, A.E. Belyaev, R.V. Konakova*

Summary

An analysis is made of the conditions for ohmic contacts realization in the case of Schottky contacts. Based on the classical notions about the mechanisms of current flow, we consider the generalized model of Schottky contact that takes into account the thermionic current of majority charge carriers and recombination current of minority charge carriers in Schottky contacts with a dielectric gap. An analysis of the results given by that model made it possible to obtain ohmicity criteria for Schottky contacts and compare the conditions for low injection level and ohmicity of Schottky contacts in the case of silicon-based contacts. It is shown that conditions for Schottky contact ohmicity do not coincide with those for *p-n* junctions.

**Введение**

До недавнего времени были известны три механизма токопрохождения, характерные для омических контактов металл-полупроводник: термоэлектронный [1,2], термополевой и туннельный [3]. В работах Гольдберга с соавторами был предложен еще один механизм, заключающийся в прохождении тока по металлическим шунтам, сопряженным с дислокациями (см., например, [4]). В работах [5-7] этот механизм был дополнен рассмотрением процессов подвода тока, протекающего в приконтактной области полупроводника, граничащей с торцами шунтов. Было показано, что вследствие крайне малых диаметров металлических шунтов на торцах шунтов возникает очень большая напряженность электрического поля, из-за чего силы зеркального изображения меняют знак изгиба зон на границе полупроводник – торец шунта с истощающего на обогащающий. Такой контакт является омическим при любых температурах, включающих и температуру жидкого гелия. В работе [5] были получены общие соотношения для удельного контактного сопротивления таких контактов. В предельном случае, когда протекающий ток ограничивается сопротивлением шунтов, справедливы выражения [3], а в другом предельном случае, когда действует ограничение тока подводом к торцам шунтов, справедливы соотношения, приведенные в [6,7].

*e-mail: sach@isp.kiev.ua

В работах [6,7] был выполнен модельный эксперимент для контактов кремний-металл с предварительно шлифованной поверхностью. Было показано, что такие контакты в отличие от контактов кремний-металл с полированной поверхностью полупроводника являются омическими. Им присуща большая плотность дислокаций, достаточная для реализации механизма токопрохождения через дислокации, сопряженные с металлическими шунтами. Отметим, что между теорией и экспериментом было получено хорошее совпадение.

Вопрос о типе контакта металл-полупроводник с шлифованной поверхностью полупроводника анализировался достаточно давно. Так, например, в ряде работ были получены омические контакты как на шлифованной поверхности кремния [8], так и на шлифованной поверхности других полупроводников [9-11]. Более того, как хорошо известно, технология получения омических контактов в структурах для силовой электроники на основе кремния обязательно содержит операцию предварительной шлифовки поверхности кремния [12,13].

В работах [14-16] были предприняты попытки объяснить получаемые результаты тем, что это контакты с высокой скоростью рекомбинации носителей заряда на границе раздела металл-полупроводник. Утверждалось, что в таких контактах концентрация носителей заряда на границе раздела между полупроводником и металлом близка к равновесной, а приконтактная область пространственного заряда (ОПЗ) отсутствует.

Работы [17-19] посвящены анализу влияния величины скорости поверхностной рекомбинации, происходящей на антизапорном тыловом контакте $p-n$ перехода, на его ВАХ. В них рассматривается поведение контактов в структурах с $p-n$ - переходами, а не в контактах Шоттки.

В настоящей работе, во-первых, получены общие соотношения для протекающих токов в контакте Шоттки с диэлектрическим зазором. Во-вторых, приведены критерии омичности контактов Шоттки. В-третьих, на основе обобщенной модели проанализировано токопрохождение в контактах Шоттки и показано, что малость уровня инъекции, когда избыточная концентрация неосновных носителей заряда (дырок) $\Delta p$ мала по сравнению с равновесной концентрацией электронов в полупроводнике $n_0$, в отличие от $p-n$ переходов не гарантирует омичности контактов Шоттки.

Предполагается, что выполнен критерий отсутствия разогрева основных носителей заряда (электронов) протекающим током $E_b < kT/ql_p$, где $E_b$ - напряженность электрического поля в объеме полупроводника, $k$ - постоянная Больцмана, $T$ - температура, $q$ - элементарный заряд, $l_p$ - длина свободного пробега электронов. Это позволяет использовать больцмановскую статистику для электронов и дырок с температурой, равной температуре решетки.

**1. Общие выражения для тока основных и неосновных носителей заряда, протекающих в контакте Шоттки с диэлектрическим зазором**

В работах [17-19], посвященных моделированию свойств антизапорного контакта к $p-n$ переходу, использовалось стоковое граничное условие для плотности тока неосновных носителей заряда (дырок) в слабо легированной $n$ - области вида

$$J_p = qS_k \Delta p(x=d), \qquad (1а)$$

где $d$ - толщина этой области.

При этом считалось, что $S_k$ (скорость поверхностной рекомбинации в плоскости контакта $x=d$) может быть как угодно велика.

Следует отметить, что соотношение (1а) нельзя использовать для нахождения плотности дырочного тока, поскольку в плоскости $x=d$, вообще говоря, существует изгиб зон, а стоковое граничное условие справедливо лишь при отсутствии изгиба зон, т.е. в плоскости, отдаленной от плоскости $x=d$ на толщину области пространственного заряда $w$ [20]. Тогда вместо (1а) в качестве граничного условия нужно использовать соотношение

$$J_p = qS_{eff} \Delta p(x = d-w), \qquad (1б)$$

где $\Delta p(x=d-w)$ - избыточная концентрация дырок в плоскости $x=d-w$, а $S_{eff}$ - эффективная скорость поверхностной рекомбинации в плоскости $x=d-w$, величина которой ограничена подводом неосновных носителей заряда к плоскости $x = d$ и не может быть как угодно большой [20].

Мы получим выражение для плотности тока неосновных носителей заряда, протекающего в контакте Шоттки с диэлектрическим зазором, исходя из более общего граничного условия, считая, что дырочный ток течет через границу раздела полупроводник-диэлектрик $x = 0$, у которой существует изгиб зон. В этом случае согласно [20] граничное условие для плотности тока неосновных носителей заряда может быть записано в виде

$$J_p = -q\,\vartheta_p \frac{V_p}{4}(p_c - p_{c0})\ ,\quad (2)$$

где $\vartheta_p$ - коэффициент прозрачности диэлектрического зазора для дырок, $V_p = \sqrt{8kT/\pi m_p}$ - средняя тепловая скорость дырок, $m_p$ - эффективная масса дырок, $p_c$ и $p_{c0}$ - неравновесная и равновесная концентрации дырок на границе полупроводника с диэлектриком в контакте металл- полупроводник.

Двукратное интегрирование уравнения непрерывности дырочного тока по координате x, перпендикулярной плоскости контакта металл – полупроводник, позволяет получить в невырожденном случае следующее выражение для концентрации неравновесных дырок в приконтактной ОПЗ:

$$p(x) = e^{-y(x)}\left(p_w - \frac{J_p}{qD_p}\int_w^x e^{y(x')}dx'\right), \quad (3)$$

где $y(x) = q\varphi(x)/kT$ - безразмерный электростатический потенциал (изгиб зон), $p_w$ - неравновесная концентрация дырок на границе ОПЗ и квазинейтрального объема в плоскости $x = w$, $D_p$ - коэффициент диффузии дырок.

Избыточная концентрация дырок на границе ОПЗ $x = w$ $\Delta p_w = p_w - p_0$ может быть найдена из уравнения баланса генерации-рекомбинации, которое в случае толстого по сравнению с длиной диффузии $L_d$ полупроводника имеет вид

$$\frac{J_p}{q} = \left(S + \frac{D_p}{L_d}\right)\Delta p_w, \quad (4)$$

где $S$ - эффективная скорость поверхностной рекомбинации в полупроводнике в плоскости $x = w$.

Использование уравнений (2) –(4) при выполнении критерия $L_d >> L_D$, где $L_D$ - длина экранирования Дебая, дает возможность получить следующее выражение для плотности дырочного тока

$$J_p = \frac{qV_{pe}p_0\left(e^{qV_s/kT}-1\right)}{1+\dfrac{V_{pe}}{S+D_p/L_d}}. \qquad (5)$$

Здесь $V_{pe} = \vartheta_p V_p e^{-y_c}$ - эффективная скорость эмиссии дырок из полупроводника в металл, $p_0$ - равновесная концентрация дырок в нейтральном объеме, $y_c = y_{c0} + \dfrac{qV_s}{kT}$ - безразмерный неравновесный изгиб зон на границе раздела $x=0$, $V_s$ - часть приложенного к диодной структуре напряжения $V$, падающая в полупроводнике.

Если выполнен критерий

$$V_{pe} >> S + \frac{D_p}{L_d}, \qquad (6)$$

то величина плотности дырочного тока равна

$$J_p = qp_0\left(S + \frac{D_p}{L_d}\right)\left(e^{qV_s/kT}-1\right) \qquad (7)$$

и соответствует плотности тока, протекающего в асимметричном $p-n$ переходе.

Аналогичным образом может быть рассчитана плотность протекающего электронного тока $J_n$. Если механизм протекания электронного тока термоэлектронный, а полупроводник не вырожден, то согласно [20]

$$J_n = \frac{q}{4}\vartheta_n V_n(n_c - n_{c0}), \qquad (8)$$

где $\vartheta_n$ - коэффициент прозрачности диэлектрического зазора для электронов, $V_n = \sqrt{8kT/\pi m_n}$ - средняя тепловая скорость электронов, $m_n$ - эффективная масса для электронов, $n_c$ и $n_{c0}$ - неравновесная и равновесная концентрации электронов на границе $x=0$.

Двукратное интегрирование уравнения непрерывности электронного тока по координате $x$, перпендикулярной плоскости контакта металл – полупроводник, дает следующее выражение для концентрации электронов в ОПЗ:

$$n(x) = e^{y(x)}\left[n_0 + \int_w^x \frac{e^{-y(x')}}{D_n}\frac{J_{nc}}{q}dx'\right]. \qquad (9)$$

Здесь $D_n$ - коэффициент диффузии электронов.

Введем величину

$$V_{nr} = D_n e^{-y_c} / \int_0^w e^{-y(x)} dx \ , \qquad (10)$$

имеющую физический смысл скорости прохождения электронов через ОПЗ полупроводника. В случае, когда во всей ОПЗ полупроводника электростатический потенциал изменяется по закону Шоттки, величина $V_{nr}$ равна дрейфовой скорости электронов $V_{nr} = \mu_n E_c$, где $\mu_n$ - подвижность электронов, а $E_c$ - напряженность электрического поля в плоскости контакта.

Подставив (9) в (8) (при x=0), можно получить связь между значением концентрации электронов в нейтральном объеме и при x=0. Окончательное выражение для плотности электронного тока, текущего через контакт, с учетом (8) имеет вид:

$$J_n = \frac{q}{4} \vartheta_n V_n n_0 \left( \frac{e^{y_c} - e^{y_{c0}}}{1 + \vartheta_n V_n / 4V_{nr}} \right), \qquad (11)$$

где $y_c = y_{c0} + \dfrac{qV_s}{kT}$.

Если выполнен критерий $\dfrac{1}{4}\vartheta_n V_n \ll V_{nr}$, то для электронов выполняется диодная теория, а выражение для плотности протекающего электронного тока упрощается и принимает такой вид

$$J_n = \frac{q}{4} \vartheta_n V_n n_0 e^{y_{c0}} \left( e^{qV_s/kT} - 1 \right). \qquad (12)$$

Проводя дальнейший анализ, значения $\vartheta_p$ и $\vartheta_n$ будем рассматривать как параметры задачи. При построении рисунков будем также считать, что $\vartheta_p = \vartheta_n$, а $V_s \cong V$. Последнее условие означает, что изменением падения напряжения в диэлектрике при приложении напряжения $V$ можно пренебречь по сравнению с изменением падения напряжения в приконтактной области полупроводника.

В работах [14-16] безотносительно к тому, о каких структурах, $p-n$ - переходах или диодах Шоттки идет речь, говорится, что «рекомбинационные» омические контакты - это контакты с высокой скоростью рекомбинации носителей заряда на границе раздела металл-полупроводник. Не возражая против того, что так называемый «рекомбинационный» контакт к $p-n$ - переходу при $x = d$ может быть антизапорным т.е. омическим [17, 18], нельзя согласиться с тем, что в случае достаточно больших величин эффективной скорости поверхностной рекомбинации $S$ «рекомбинационный» контакт к диоду Шоттки также будет омическим. Величина эффективной скорости поверхностной рекомбинации $S$ (см. выражение (7)) хотя и может быть достаточно большой, однако она ограничена сверху значением тепловой скорости дырок $V_p$. По этой же причине не может быть также реализовано условие $p_c \cong p_{c0}$, когда неравновесная концентрация неосновных носителей заряда равна равновесной.

Говоря другими словами, условия омичности контактов Шоттки не соответствуют условиям омичности контактов к $p-n$ переходам. Поэтому далее мы остановимся на анализе критериев омичности к контактам Шоттки.

**2. Критерии омичности контактов Шоттки**

Рассмотрим случай, когда контакт Шоттки реализован на основе невырожденного полупроводника $n$- типа. Будем вначале считать, что выполнен критерий $J_n >> J_p$. В этом случае ток, текущий через контакт, определяется выражением вида (11) и может быть записан в таком виде

$$I_n = I_{ns}\left(\exp\left(\frac{qV}{kT}\right) - 1\right), \qquad (13)$$

где $I_{ns} = \frac{q}{4} A \vartheta_n V_n n_{c0} /(1 + \vartheta_n V_n / 4V_{nr})$, $A$ - площадь контакта, $n_{c0} = n_0 \exp\left(-\frac{q\varphi_b}{kT}\right)$, $\varphi_b$ - электростатический потенциал на границе полупроводник-диэлектрик, отсчитанный от дна зоны проводимости полупроводника.

Согласно [3] контактное сопротивление равно

$$R_c = \left(\frac{dI_n}{dV}\right)^{-1}_{V=0}. \qquad (14)$$

Подстановка (13) в (14) дает

$$R_c = \frac{kT}{q}\frac{1}{I_{ns}}. \qquad (15)$$

Получим далее критерий омичности контакта Шоттки для рассматриваемого случая. С этой целью запишем выражение для протекающего тока с учетом объемного сопротивления $R_b = \rho d / A$, где $\rho$ - удельное сопротивление полупроводника, $d$ - толщина полупроводника, $A$ - площадь контакта:

$$I_n = I_s\left(\exp\left(\frac{q(V - IR_b)}{kT}\right) - 1\right). \qquad (16)$$

Выражение (16) можно переписать в таком виде

$$I = \frac{V}{R_b} - \frac{kT}{qR_b}\ln\left(1+\frac{I}{I_S}\right). \qquad (17)$$

При $I/I_s \ll 1$ уравнение (17) с учетом (15) приводится к виду

$$I = \frac{V}{R_b} - I\frac{R_c}{R_b}. \qquad (18)$$

Таким образом, как видно из (18), критерием омичности контакта Шоттки является неравенство $R_c \ll R_b$, т.е. в данном случае контактное сопротивление должно быть существенно меньше объемного. Этот критерий, в частности, совпадает с критерием, приведенным в монографии [3]. Такого же вида критерий омичности контакта Шоттки был получен в [21] для случая, когда протекающий в контакте Шоттки ток основных носителей заряда определяется термополевым током. Указанный случай может реализоваться при достаточно больших величинах $\varphi_b$ и (или) в области достаточно низких температур.

На рис. 1а приведены зависимости величины $R_c$ от коэффициента прозрачности диэлектрического зазора $\vartheta_n$ при использовании параметров кремния. Параметром кривых является высота барьера $\varphi_b$. Кривая 5 соответствует объемному сопротивлению $R_b$. Как видно из рисунка, критерий омичности контакта Шоттки при использованных параметрах кремния может быть выполнен лишь при $\varphi_b \leq 0.1$ В в случае достаточно большого значения прозрачности диэлектрического зазора для электронов $\vartheta_n$. Отметим, что при построении данного и следующих рисунков считалось, что $A = 1$ см$^2$, $n_0 = 10^{15}$ см$^{-3}$, $S = 10^5$ см/с, $d = 220$ мкм, а T = 300 К.

На рис. 1 б приведены зависимости величины $R_c$ от значения $\varphi_b$. Параметром кривых является величина $\vartheta_n$. Кривая 4 соответствует объемному сопротивлению. Как видно из рисунка, чем меньше величина $\vartheta_n$, тем при меньших значениях $\varphi_b$ выполняется критерий омичности контакта Шоттки.

Следует отметить, что уменьшение прозрачности зазора $\vartheta_n$ эквивалентно увеличению высоты барьера $\varphi_b$. При этом критерий омичности контакта может не выполняться даже при малых значениях $\varphi_b$ (см. кривую 2 рис. 1 а). Это следует учитывать при анализе экспериментальных ВАХ.

## 3. Анализ токов, протекающих в контактах Шоттки, с учетом тока неосновных носителей заряда

В работах [17-19] можно найти утверждение, что контакт является омическим в случае, когда реализуется малый уровень инжекции, т.е. выполнен критерий $\Delta p \ll n_0$. Покажем, что это утверждение неприменимо к контактам Шоттки.

В случае, когда можно пренебречь падением приложенного напряжения в диэлектрике, выражение для избыточной концентрации дырок $\Delta p$ имеет следующий вид

$$\Delta p = \frac{p_0 V_{pe}}{V_{pe} + S + \dfrac{D_p}{L_d}} \left[ \exp\left(\frac{q(V - IR_b)}{kT}\right) - 1 \right]. \qquad (19)$$

Отметим, что в данном случае $\Delta p = \Delta p_w$, что следует из условия постоянства квазиуровня Ферми для дырок в приконтактной ОПЗ.

Рассмотрим далее случай, когда полный протекающий ток в контакте Шоттки $I$ равен сумме электронного и дырочного токов, т.е. $I = I_n + I_p$, где величины $I_n = AJ_n$ и $I_p = AJ_p$, а значения $J_n$ и $J_p$ определяются выражениями (11) и (5) соответственно. Как и раньше, будем считать, что падением приложенного напряжения в диэлектрике можно пренебречь по сравнению с падением напряжения в полупроводнике. Тогда полный ток, протекающий в контакте Шоттки, может быть записан как

$$I = \left( I_{ns} + \frac{qAp_0 V_{pe}}{1 + \dfrac{V_{pe}}{S + D_p/L_d}} \right) \left( \exp\left(\frac{q(V - IR_b)}{kT}\right) - 1 \right). \qquad (20)$$

На рис. 2а приведены зависимости электронного тока $I_n$ и дырочного тока $I_p$, пропорциональных соответственно первому и второму слагаемому в круглой скобке (20), от величины коэффициента прозрачности диэлектрического зазора для электронов $\vartheta_n$ и равного ему коэффициента прозрачности диэлектрического зазора для дырок $\vartheta_p$. Параметром кривых является величина $\varphi_b$. Как видно из рисунка, при использованных при построении рисунка значениях $\varphi_b$ (за исключением случая, когда $\varphi_b = 0.2$ В) величина дырочного тока постоянна и не зависит ни от коэффициента прозрачности диэлектрического зазора, ни от величины $\varphi_b$. Однако по мере уменьшения $\vartheta_n$ сильно уменьшается значение $I_n$. За счет этого при малых значениях $\vartheta_n$ ток дырок может стать больше, чем ток электронов. В случае, когда $I_p > I_n$, контакт Шоттки ведет себя как $p-n$ переход, а при $I_p < I_n$ - как классический контакт Шоттки. Чем меньше величина $\vartheta_n$, тем при меньших значениях $\varphi_b$ реализуется условие $I_p > I_n$. Следует отметить, что в случае, когда $\vartheta_n = 1$, условие $I_p \geq I_n$ при использованных значениях параметров реализуется, если $\varphi_b \geq 0.71$ В.

На рис. 2б приведены зависимости электронного тока $I_n$ и дырочного тока $I_p$ от величины $\varphi_b$. Параметром кривых для $I_n$ является значение $\vartheta_n$, а для $I_p$ - величина $\vartheta_p$, которые при построении рисунка полагались равными. При использованных значениях $\vartheta_p$ величина дырочного тока $I_p$ не зависит от значений $\vartheta_p$ и $\varphi_b$, начиная со значений $\varphi_b > 0.2$ В. При $\varphi_b \leq 0.2$ В величина $I_p$ уменьшается по мере уменьшения значения $\varphi_b$. Это уменьшение связано с нарушением критерия (6), вследствие чего контакт Шоттки не пропускает весь инжектированный дырочный ток.

На рис. 3 с использованием формулы (19) построены зависимости избыточной концентрации дырок $\Delta p$ от приложенного напряжения $V$. Параметром кривых является значение $\varphi_b$.

И, наконец, на рис. 4 с использованием (20) приведены зависимости суммарного тока, протекающего в контакте Шоттки, от приложенного напряжения. Как и на рис. 3, параметром кривых является величина $\varphi_b$.

Отметим, что как рис. 3, так и рис. 4 построены для случая, когда $\vartheta_p = \vartheta_n = 1$. Как видно из рис. 3, при использованных параметрах для всех значений $\varphi_b$ выполняется условие $\Delta p < n_0$, т.е. уровень инжекции является малым. В то же время, как видно из рис. 4, при $\varphi_b \geq 0.3$ В для всех кривых реализуется четко выраженный участок экспоненциальной зависимости протекающего тока от приложенного напряжения. Это свидетельствует о том, что в случае контактов Шоттки реализация малого уровня инжекции в отличие от случаев, рассмотренных в работах [17-19], не является достаточной для реализации условия омичности контакта.

**4. Обсуждение полученных результатов**

Ток, текущий через контакт металл - полупроводник, зависит от эффективной скорости поверхностной рекомбинации лишь в случае, когда ток неосновных носителей заряда превышает ток основных носителей заряда, а для этого контакт Шоттки должен действовать как $p - n$ переход. Это возможно лишь при больших величинах барьера.

Однако в этом случае контактное сопротивление становится большим и условие омичности контакта $R_c \ll R_b$ не выполняется. При этом контакт будет неомическим.

Хотя в данном случае неравновесная концентрация носителей заряда на контакте близка к равновесной, вопреки утверждениям работ [14-16] изгиб зон на границе раздела полупроводник- диэлектрик не равен нулю. Как показывает анализ, при приложении смещения $V$ в прямом направлении приложенное напряжение вначале падает на приконтактной ОПЗ. При этом ВАХ контакта имеет выпрямляющий характер. Естественно, что при $V \leq \varphi_b$ область пространственного заряда будет существовать.

Что касается сильного влияния большого значения скорости поверхностной рекомбинации $S$, приводящего к уменьшению величины $\Delta p$, то оно не так уж и велико. Если учесть, что значение $S \leq V_p \approx 10^7$ см/с [20], то, как показывают оценки, даже при $S \approx 10^7$ см/с величина $\Delta p$ при условии $J_p = J_n$ и при $V = 0.6$ В примерно равна $10^{12}$ см$^{-3}$, в то время как значение $p_0 \approx 10^5$ см$^{-3}$. Т.о., в данном случае избыточная концентрация дырок в плоскости $x = w$ на семь порядков превышает равновесную концентрацию. Пересчет величины скорости рекомбинации дырок в плоскости $x = w$, равной $S$, на значение в плоскости контакта дает величину $V_p / 4$ (при $\vartheta_p = 1$). Поэтому устремлять величину $S_{eff}$, определяемую соотношением вида (1б), к бесконечности нельзя.

Отметим, что термоэлектронный ток основных носителей заряда не зависит от рекомбинационных характеристик, в частности, от поверхностной рекомбинации. Отсутствует подобная зависимость и для термополевого, а также для туннельного тока. Указанная зависимость существует лишь в случае p-n переходов.

При анализе мы ограничились рассмотрением только термоэлектронного механизма токопрохождения основных носителей заряда в контактах Шоттки. Можно отметить, что ситуация качественно не изменится, если механизм токопрохождения основных носителей заряда в контактах Шоттки термополевой. Качественное изменение происходит в случае, когда доминирует туннельный механизм токопереноса в контакте Шоттки. В этом случае объем полупроводника вырожден и чем больше уровень легирования, тем больше степень вырождения, поэтому эффективная высота барьера, отсчитанная от уровня Ферми в полупроводнике, с ростом уровня легирования уменьшается. Толщина барьера при этом также уменьшается, что приводит к увеличению его прозрачности. Соответственно уменьшается и контактное сопротивление. Поэтому контакт может работать как омический.

Как следует из полученных результатов, условие омичности контакта Шоттки $R_c << R_b$ выполняется тем лучше, чем больше величина коэффициента прозрачности диэлектрического зазора для электронов $\vartheta_n$. Поскольку $\vartheta_n \propto \exp(-\alpha d_d)$ [20], где $\alpha$ - постоянная, а $d_d$ - толщина диэлектрического зазора, то для увеличения $\vartheta_n$ следует реализовать минимальные значения $d_d$. Величина $\alpha$, в свою очередь, зависит от ширины запрещенной зоны диэлектрика $E_d$. Она возрастает с ее увеличением и уменьшается с уменьшением. В случае контактов Шоттки на основе кремния величина $E_d$ определяется степенью стехиометрии окисла $SiO_x$. Чем меньше величина $x$ по сравнению с двойкой, тем меньше значение $E_d$ и, соответственно, меньше величина $\alpha$.

Обсудим далее приближение $\vartheta_n = \vartheta_p$, использованное при анализе. Вообще говоря, указанное равенство не имеет места. Однако, в тех случаях, когда высота барьера $\varphi_b$ достаточно большая ($> 0.3$ В), величина дырочного тока при использованных для расчета параметрах определяется выражением (7) и не зависит от коэффициента прозрачности диэлектрического зазора для дырок $\vartheta_p$ при его изменении на несколько порядков по величине. Это, тем более, имеет место в тех случаях, когда выполнено неравенство $I_p > I_n$, т.е., когда реализуется $p-n$ переход, поскольку при этом значение $\varphi_b$ еще больше. Однако в случае, когда критерий (6) нарушен, протекающий дырочный ток будет пропорционален величине $\vartheta_p$.

При меньших значениях $\varphi_b$ реализуется контакт Шоттки, поскольку $I_n >> I_p$. В этом случае протекающий электронный ток $I_n$ пропорционален величине $\vartheta_n$. Ошибка, получаемая при замене $\vartheta_p$ на $\vartheta_n$ в случае расчета величины $I_p$, хотя и приведет неправильному значению $I_p$, однако практически не скажется на суммарной величине протекающего тока $I$.

**5. Заключение**

Т.о. из проведенного выше анализа, а также с учетом результатов работ [4-7] можно сделать следующие выводы.

1. Контакт Шоттки является омическим при выполнении неравенства $R_c << R_b$. Чем меньше коэффициент прозрачности для основных носителей заряда $\vartheta_n$ в контакте Шоттки, тем хуже выполняется условие омичности контакта. Как показали проведенные расчеты, контакты Шоттки на основе кремния при температурах $\geq 300$ К являются омическими, если высота барьера, отсчитанная от дна зоны проводимости, составляет по величине $\leq 0.1$ В, а значение $\vartheta_n$ близко к 1.

2. Условие малости уровня инжекции, когда $\Delta p << n_0$, не гарантирует омичности контакта Шоттки.

3. «Рекомбинационные» контакты Шоттки, т.е. контакты, в которых преобладает ток неосновных носителей заряда, при любых достижимых значениях скорости рекомбинации $S$ являются выпрямляющими контактами.

4. Объяснение того, почему после шлифовки полупроводника образуется омический контакт, лежит за пределами рассмотренной выше модели. Это может быть понято лишь на основе представлений, развитых в работах [4-7].

Подписи к рисункам.

Рис. 1. Зависимости контактного сопротивления $R_c$ от коэффициента прозрачности диэлектрического зазора для электронов $\vartheta_n$ (рис 1а) и от высоты барьера $\varphi_b$ (рис 1б). Использованные при построении рисунка параметры: рис. 1 а - $\varphi_b$, В: 1- 0.01, 2 – 0.1, 3 – 0.3, 4 – 0.5; рис. 1 б - $\upsilon_n$: 1 – 1, 2 – $10^{-1}$, 3 - $10^{-2}$.

Рис. 2. Зависимости токов для электронов (кривые 1 -3) и для дырок (кривые 4-6) от коэффициента прозрачности диэлектрического зазора $\vartheta_n$ (рис 2а) и от высоты барьера $\varphi_b$ (рис. 2б). Использованные при построении рисунка параметры: рис. 2 а - $\varphi_b$, В: 1,4- 0.2, 2,5 – 0.4, 3,6 – 0.6; рис. 2 б - $\upsilon_n$ ($\vartheta_p$): 1,4 – $10^{-4}$, 2,5 – $10^{-2}$, 3,6 - 1.

Рис. 3. Зависимости избыточной концентрации дырок от приложенного напряжения. Использованные при построении рисунка значения $\varphi_b$, В: 1 – 0.1; 2 – 0.3; 3 – 0.5; 4 – 0.7; 5 – 0.9. Кривая 6 – значение $\Delta n$, равное 0.1 $n_0$.

Рис. 4. Зависимости протекающего тока от приложенного напряжения. Использованные при построении рисунка значения $\varphi_b$, В: 1 – 0.1; 2 – 0.3; 3 – 0.5; 4 – 0.7; 5 – 0.9.

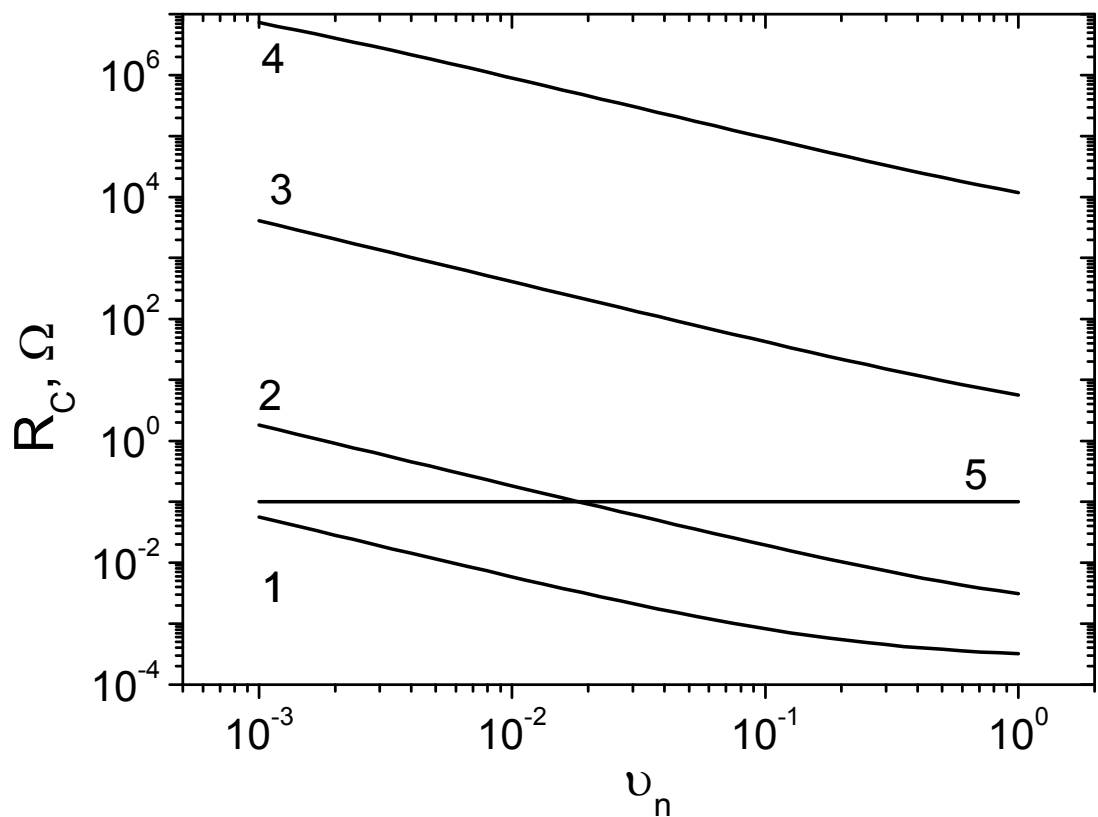

Рис 1а к статье А.В. Саченко, А.Е. Беляева и Р.В. Конаковой "К вопросу об омичности контактов Шоттки"

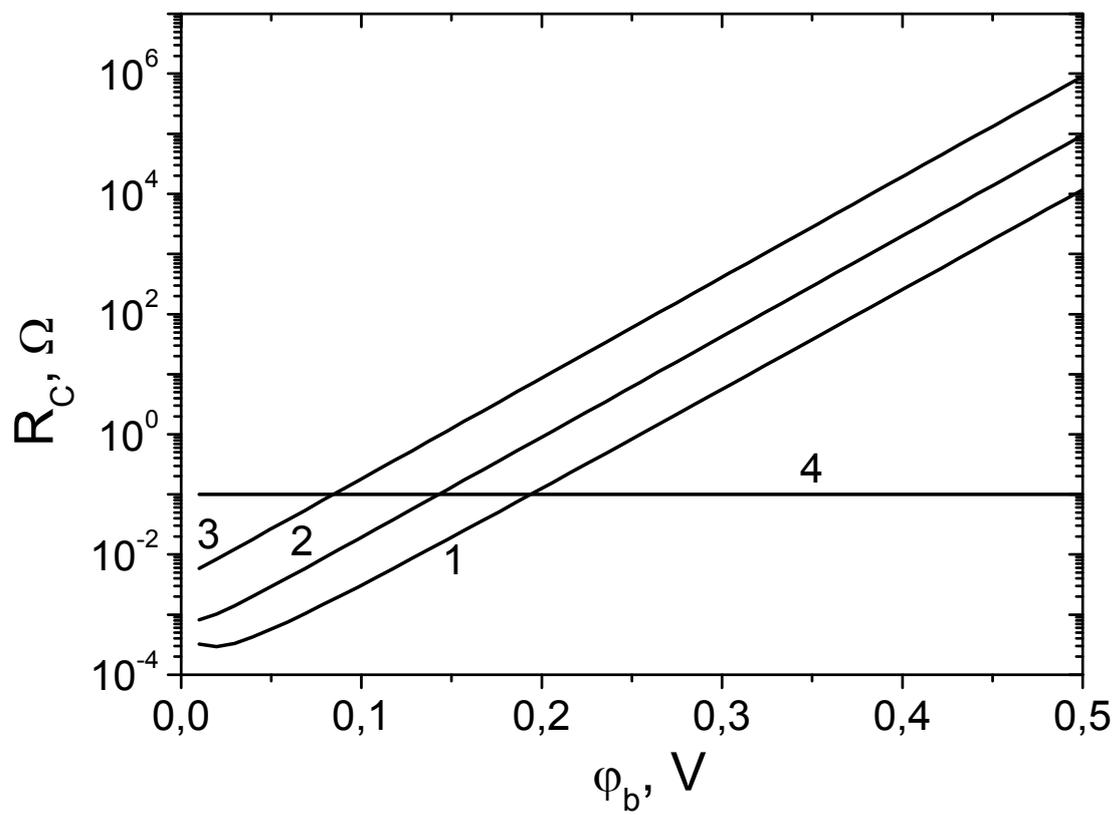

Рис 1б к статье А.В. Саченко, А.Е. Беляева и Р.В. Конаковой "К вопросу об омичности контактов Шоттки"

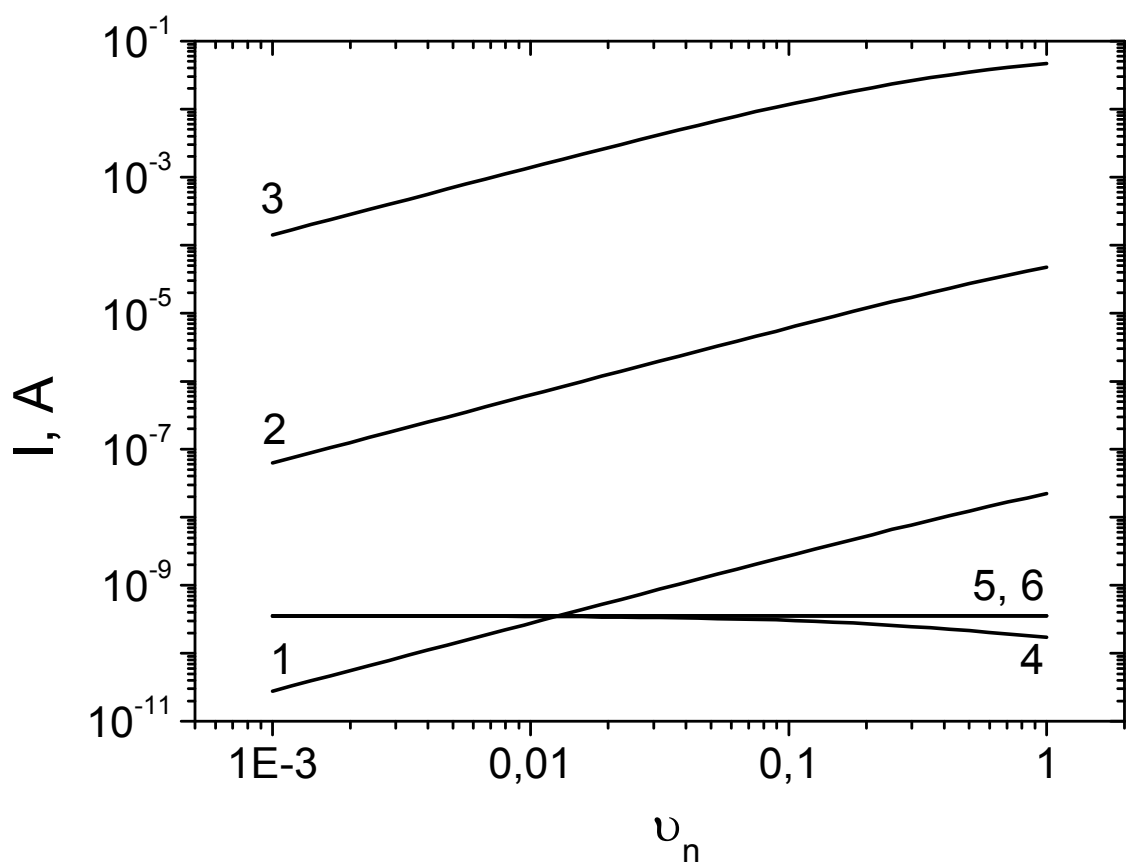

Рис 2а к статье А.В. Саченко, А.Е. Беляева и Р.В. Конаковой "К вопросу об омичности контактов Шоттки"

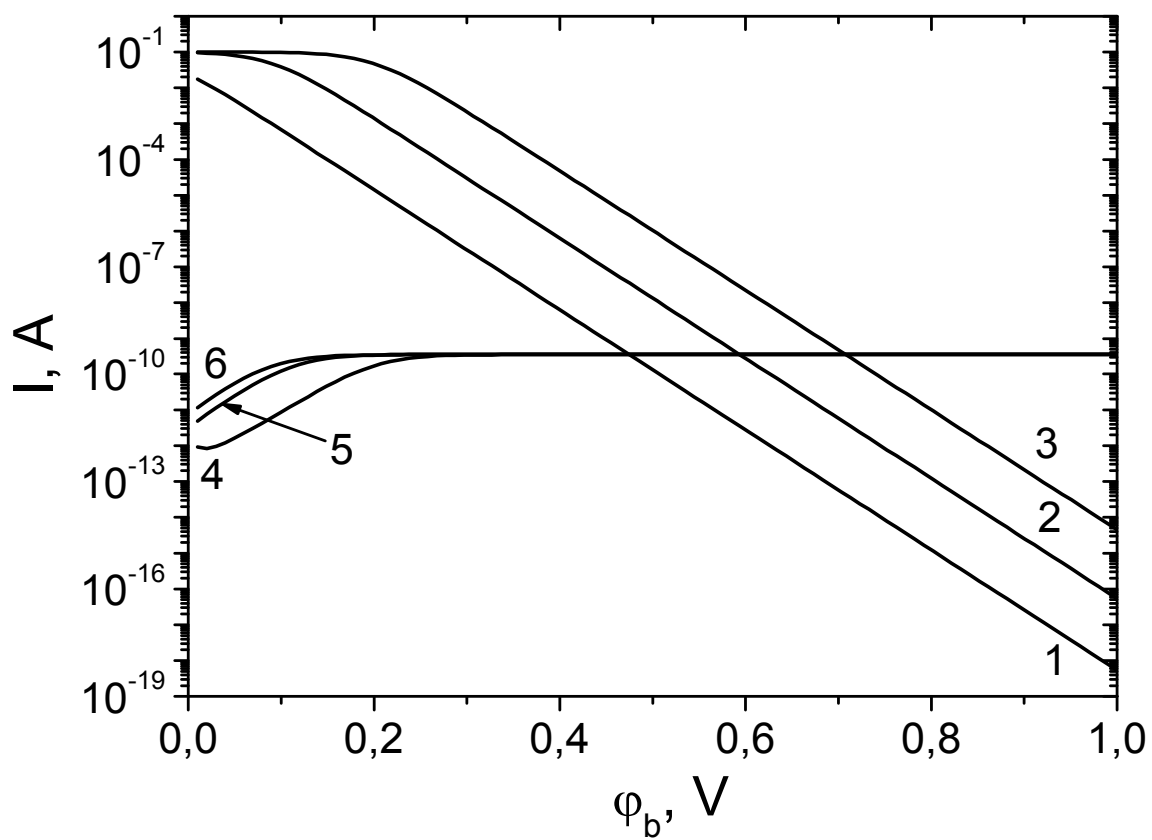

Рис 2б к статье А.В. Саченко, А.Е. Беляева и Р.В. Конаковой "К вопросу об омичности контактов Шоттки"

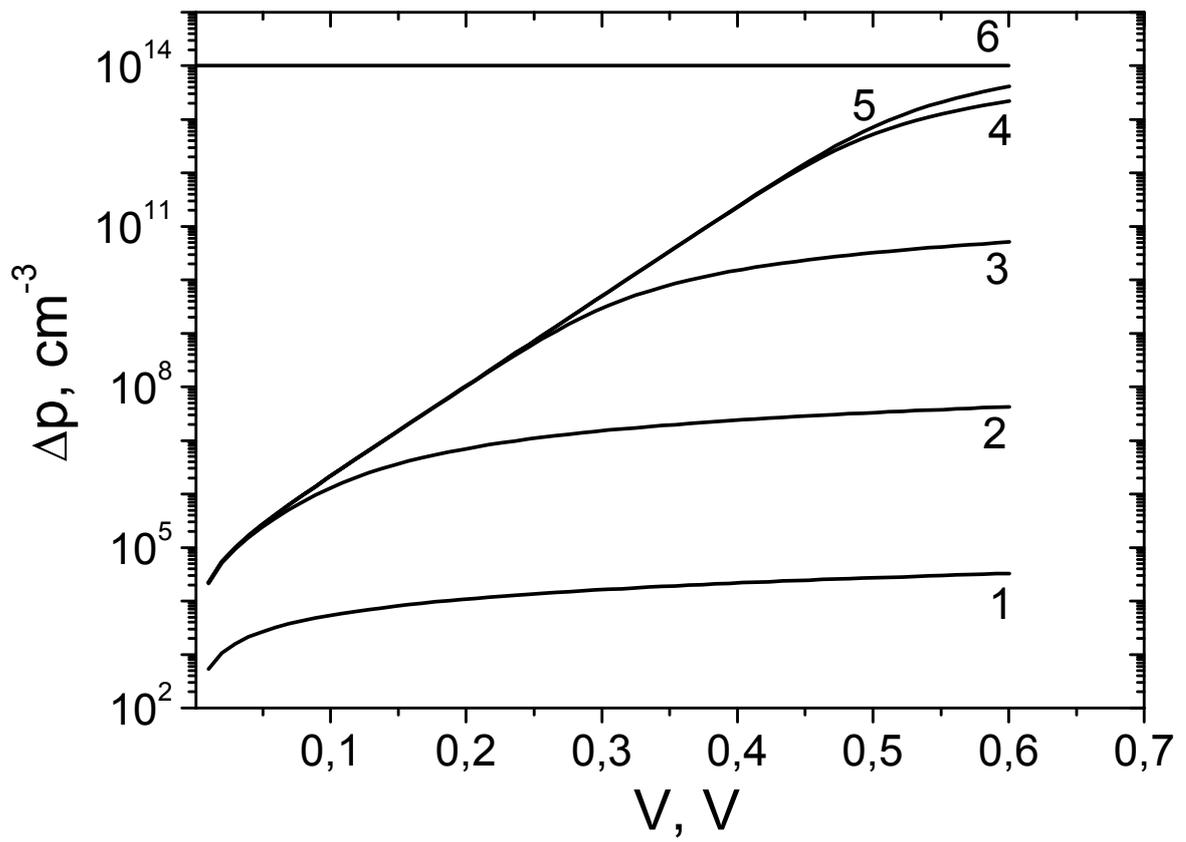

Рис 3 к статье А.В. Саченко, А.Е. Беляева и Р.В. Конаковой "К вопросу об омичности контактов Шоттки"

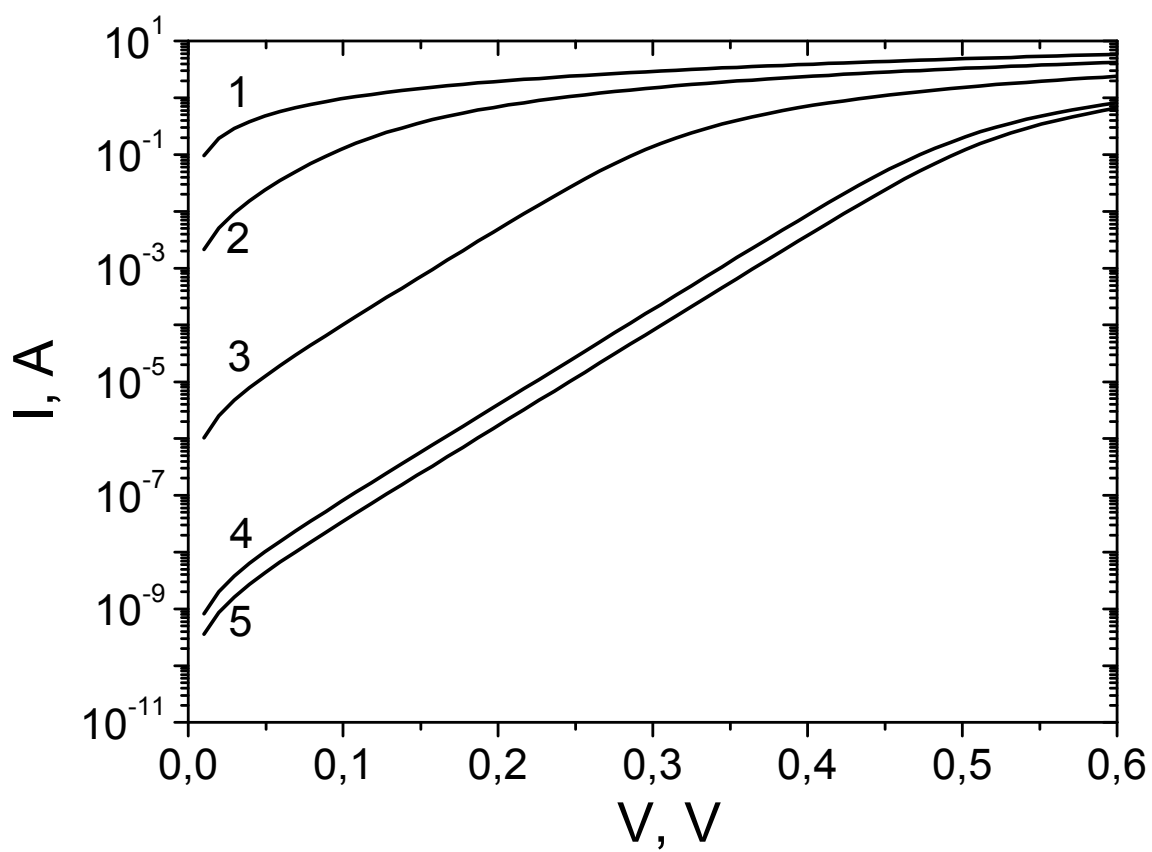

Рис 4 к статье А.В. Саченко, А.Е. Беляева и Р.В. Конаковой "К вопросу об омичности контактов Шоттки"